# New Evidence for DM-like Anomalies in Neutron Multiplicity Spectra

**W. H. Trzaska,** [a, b,*] **A. Barzilov,** [c] **T. Enqvist,** [a] **K. Jedrzejczak,** [d] **M. Kasztelan,** [e] **P. Kuusiniemi,** [a] **K. K. Loo,** [a] **J. Orzechowski,** [e] **M. Słupecki,** [f] **J. Szabelski,** [g] **and T. E. Ward** [h, i]

[a] *Department of Physics, University of Jyväskylä, Finland*

[b] *Helsinki Institute of Physics (HIP), University of Helsinki, Finland*

[c] *University of Nevada Las Vegas, Department of Mechanical Engineering, USA*

[d] *Jagiellonian University, Kraków, Poland*

[e] *National Centre for Nuclear Research (NCBJ), Poland*

[f] *CERN, Switzerland*

[g] *Stefan Batory Academy of Applied Sciences, Skierniewice, Poland*

[h] *Office of Nuclear Energy, DOE, USA*

[i] *TechSource, Santa Fe, NM, USA*

E-mail: trzaska@jyu.fi

Subterrestrial neutron spectra show weak but consistent anomalies at multiplicities ~100 and above. The origin of the excess events remains ambiguous, but, in principle, it could be a signature of Dark Matter WIMP annihilation-like interaction with a massive Pb target. However, since the results of the available measurements are below the 5-sigma discovery level, and the observed anomalous structures are on a significant muon-induced background, an independent verification at even greater depth is needed. For that purpose, we have launched NEMESIS 1.4 – a new dedicated experiment consisting of an 1134 kg Pb target and 14 helium-3 detectors with PE moderators and a fully digital readout. NEMESIS 1.4 has been taking data at the deepest level (1.4 km, 4000 m.w.e.) of the Pyhäsalmi mine, Finland, since November 2022. We describe the idea behind the new setup, compare the first results with the previous data and Monte Carlo simulations, and give the outlook for further research. If the existence of the anomalies is unambiguously confirmed and the model interpretation positively verified, this will be the first Indirect Detection of Dark Matter in the laboratory.

*Speaker

1. **Introduction**

The significance of Dark Matter (DM) studies hardly needs an explanation. Over the past decades, we have amassed impressive astronomical evidence not only supporting DM's existence but also determining its share in the mass-energy balance of the Universe. And yet, Dark Matter persistently eludes laboratory detection. The mainstream of DM searches concentrates on the Direct Detection of the hypothetical particles known as Weakly Interacting Massive Particles or WIMPs. Several high-profile experiments [1] look for the tiny signals expected from particles recoiling after a WIMP scattering. So far, no such events have been identified. However, if WIMPs exist, they must also decay or annihilate. Searches for annihilation or decay signs are called Indirect Detection.

Experiments in the Pyhäsalmi mine are the first dedicated attempt at indirect terrestrial DM detection. Our measurements consistently see small but statistically significant excess events in neutron multiplicity spectra from Pb targets. These excess events appear to have peak-like structures, discernible in long-exposure spectra. Such structures cannot originate from muon-induced spallation alone. It is, therefore, possible that the missing contribution comes from WIMP interactions [2].

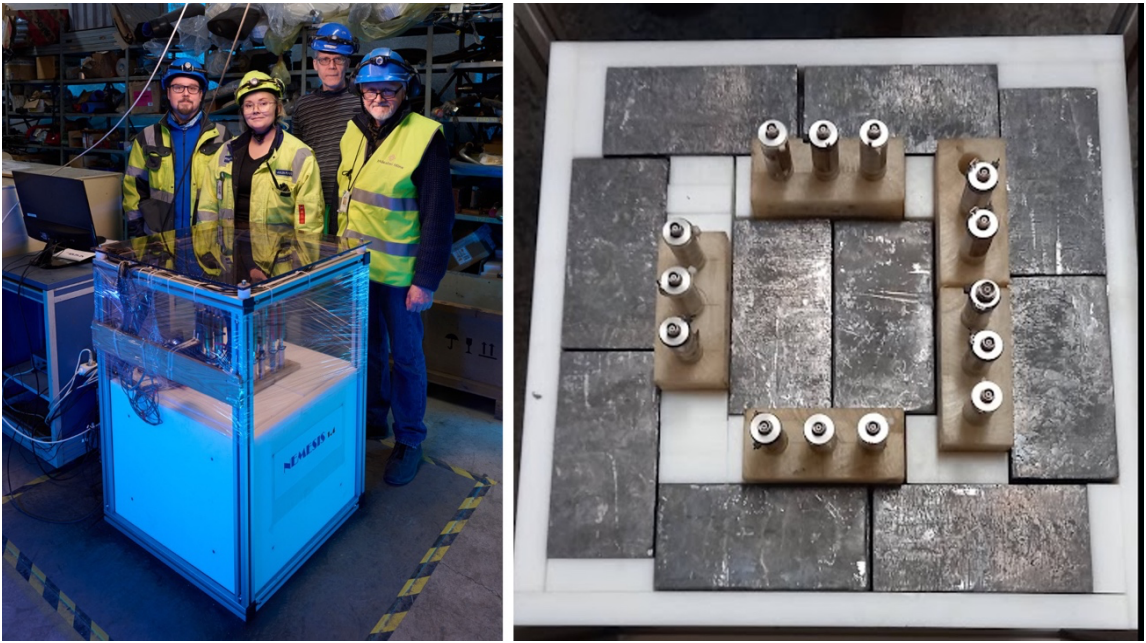

**Figure 1.** (Left) Photo of the fully assembled NEMESIS 1.4 setup. (Right) Configuration of the target (100 Pb bricks in 10 stacks of 10 bricks each) and fourteen helium-3 counters in PE casting. The picture was taken before adding the electronics, the external PE layer, and protective covers.

However, Dark Matter annihilation is not the only process leading to significant neutron emissions. They are also produced by high-energy cosmic-ray muons passing through matter. To reduce that background, we also measured 1.4 km below the surface (4000 m.w.e.). We now have preliminary results from the measurements at 0.1, 0.2, 0.4, and 1.4 km underground.

## 2.  NEMESIS 1.4 experiment

The first NEMESIS experiment was located ~0.1 km underground and operated from November 2019 to November 2022. In November 2022, the setup was upgraded and installed ~1.4 km underground. The name of the new experiment is NEMESIS 1.4. The core name NEMESIS stands now for "NEutron MEasurementS In Sub-terrestrial locations". It is the essence of our measurements. We look for signs of WIMP annihilation in massive targets placed underground. A tell-tale feature of such a violent event would be a neutron burst. Our experiments look for such neutrons.

Figure 1 shows photos of the NEMESIS 1.4 setup. The target is made of 100 standard-size Pb bricks (5 x 10 x 20 cm$^3$), arranged in 10 piles of 10 bricks each, as visible in the right photo. The total weight of Pb is 1,134 kg. Also visible are 14 helium-3 neutron counters located between the Pb bricks. The counters are clustered in groups of three and placed inside a 15 x 6.4 x 55 cm$^3$ polyethylene (PE) casting. The available space between the bricks and detectors is filled with PE. The outer shielding is made of 3 cm thick PE plates. The total PE weight is 130 kg. It serves as a neutron moderator, reflector, and shielding.

Each neutron counter is 50 cm long and has a built-in preamplifier and a digitiser sampling the signal for 0.3 ms before and 1.7 ms after the trigger. The trigger is generated whenever any detector registers a valid signal. The sampling width is sufficient to account for variations in neutron thermalisation. The sampling step and the software allow the detection of multiple neutrons in the same counter if they are spaced by more than 2-3 μs. Data-taking and operation are controlled remotely. The extraction of neutron multiplicity from the digitised waveforms is described in [3].

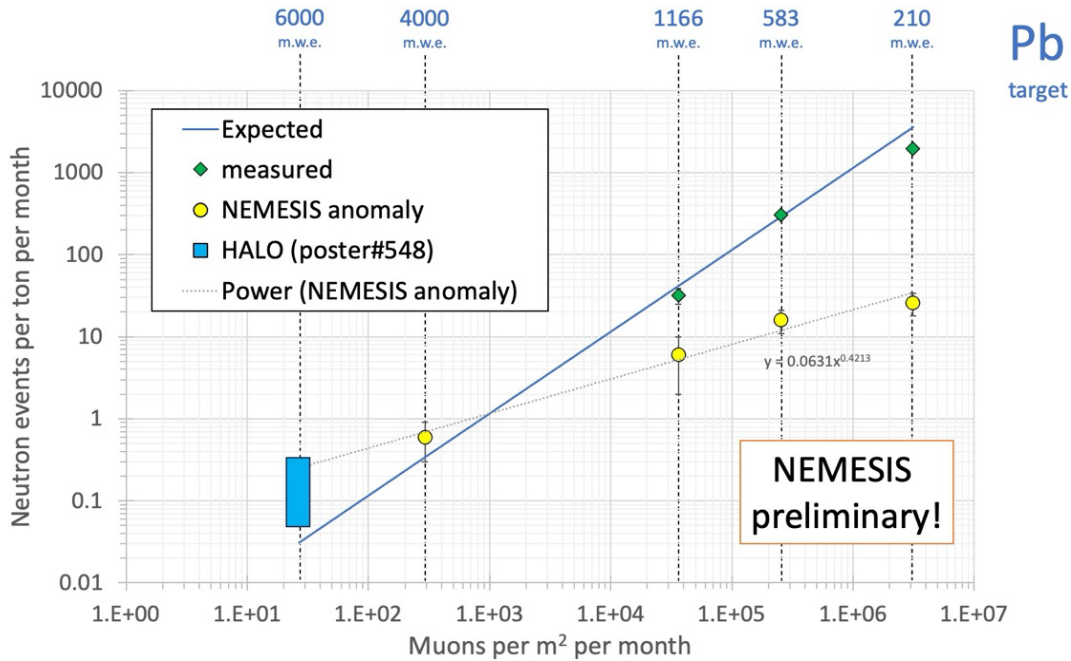

**Figure 2.** Compilation of preliminary results taken at the different depths. Additional explanation is provided in chapter 3.

## 3. Preliminary Results

The first NEMESIS 0.1 results were presented at ICRC 2021 [4], TAUP 2021 [5], VCI 2022 [6], and NDM 2022 [7]. Since then, we have conducted detailed Monte Carlo studies [8] for the NMDS setup used for measurements at 0.2 km (583 m.w.e) and 0.4 km (1166 m.w.e.). The simulations confirmed the setup's detection efficiency and the shape of cosmic-ray-induced neutron spectra. We have also analysed the total (ungated) NEMESIS 0.1 neutron multiplicity spectrum and, for the first time, present half-a-year data from NEMESIS 1.4 – the measurement at 1.4 km underground.

The preliminary results are summarised in Fig. 2. It shows the measured number of neutron events per month per 1000 kg of the target mass as a function of the depth-related muon flux. This compilation includes all experiments carried out in the Pyhäsalmi mine over the past two decades. The overburden (in m.w.e.) at each site is also indicated. The muon flux is known from previous measurements [9]. The green diamond-shaped points show the muon-induced neutron events measured at shallower depths. Such events should scale linearly with the muon flux, as illustrated with the fitted blue solid line, extrapolated to the lower fluxes. The NEMESIS anomalies are marked as yellow points with error bars. The yellow point at 4000 m.w.e. represents the first six months of NEMESIS 1.4 operation. The slope of the grey dotted line fitted to the anomalies differs from the blue line. Hence, although we see some depth-related dependence of the anomalies, they cannot be caused exclusively by cosmic ray-induced muons. The origin of this behaviour is unclear, but at this stage, we preliminarily attribute it to a combination of muon-based and WIMP-based sources [2]. The work is in progress. More experimental data and thorough simulations are needed. The preliminary HALO result [10] is the large blue rectangle at 6000 m.w.e. The HALO collaboration operates a 79-ton experiment at the deepest laboratory in the world. Unfortunately, analysis and interpretation of their results will be challenging because of the specific geometry and position dependence in neutron detection efficiency. HALO reports no anomalies in the spectrum, yet the preliminary total of high-multiplicity neutron events seems to follow the NEMESIS trend. We eagerly wait for the final HALO results.

## 4. Summary and Outlook

The NEMESIS project, incorporating NMDS, NEMESIS 0.1, and the currently operating NEMESIS 1.4 setup, is an important niche experiment. Focusing on signals from WIMP annihilation (Indirect Detection) addresses the blind spot in the mainstream terrestrial Dark Matter searches looking only for recoils (Direct Detection). NEMESIS sensitivity for Spin Independent WIMP-nucleon annihilation cross section is ~$10^{-45}$ cm$^2$ for WIMP masses between 0.1 and 10 GeV/c$^2$ [7], thus probing previously uncharted areas. For Spin Dependent interactions, the corresponding limits will be four orders of magnitude higher [3]. The low-budget, low-maintenance NEMESIS 1.4 setup, commissioned in November 2022, must collect data for a couple more years to bring the expected results. Apart from repairs, inspections or modifications, no human presence in the mine is required. The system operates remotely. In the future, it would be desirable to significantly increase the number of neutron detectors to bring the efficiency to a 20-40% range. Adding a muon tracker with time-of-flight capabilities would help separate possible target emissions of energetic leptons from the top-down muon tracks.

**Acknowledgements**

We express our gratitude to Callio Lab [https://calliolab.com] for the availability of the underground locations and infrastructure of the Pyhäsalmi mine. Financial support from the Helsinki Institute of Physics, TechSource and Wihuri Foundation is gratefully acknowledged. Our special thanks go to the University of Oulu, Kerttu Saalasti Institute team of Dr Ossi Kotavaara, Mr Jari Joutsenvaara, and Ms Julia Puputti for help and support in setting up and running the measurements.